\documentclass[12pt]{ORSNZ}
\usepackage{amssymb, vmargin,  ORSNZ}
\setpapersize{A4}
\setmargnohfrb{30mm}{20mm}{30mm}{20mm} 

\usepackage{palatino,epsfig,amsmath, amssymb, amsfonts, color, graphics, multirow} 
\usepackage{url} 
\usepackage[hang,small,bf]{caption}

\title{Modeling Repairs of Systems with a Bathtub-Shaped Failure Rate Function} 
\author{Sima Varnosafaderani \\
        School of Mathematics, Statistics and Operations Research \\
        Victoria University of Wellington \\
        New Zealand \\
        sima\@@msor.vuw.ac.nz \vspace{5mm}\\ 
        Stefanka Chukova\\
        School of Mathematics, Statistics and Operations Research \\
        Victoria University of Wellington \\
        New Zealand \\
        stefanka\@@msor.vuw.ac.nz \vspace{5mm}\\
}
\date{} 
\begin{document}

\maketitle
\pagestyle{empty} 
\thispagestyle{empty}


\begin{abstract}
Most of the reliability literature on modeling the effect of repairs on systems
assumes the failure rate functions are monotonically increasing. For systems
with non-monotonic failure rate functions, most models deal with minimal
repairs (which do not affect the working condition of the system) or
replacements (which return the working condition to that of a new and identical
system). We explore a new approach to model repairs of a system with a
non-monotonic failure rate function; in particular, we consider systems with a
bathtub-shaped failure rate function. We propose a repair model specified in
terms of modifications to the virtual age function of the system, while
preserving the usual definitions of the types of repair (minimal, imperfect and
perfect repairs) and distinguishing between perfect repair and replacement. In
addition, we provide a numerical illustration of the proposed repair model.
\end{abstract}

\section{Introduction}

Most engineered systems -- defined as an arrangement of components that together
perform an identified (and predefined) set of functions -- are susceptible to
failures, and require some form of rectification in order to return to a
functioning condition. Most rectifications have an effect on the probability
and number of future failures of the system over a given period of time.

The sequence of numbers of failures of the system in time, i.e. the failure
process, is modeled as a stochastic counting process, assuming that there can
be at most one failure in an infinitesimally small interval of time. When the
rectification action following each failure is immediate and instantaneous,
this process can also be described as the sequence of times to failure (or
consecutive system's lifetimes).

A counting process is completely described by its conditional intensity
function, and therefore, rectifications are usually defined in terms of their
effect on the conditional intensity function of the failure process. The
initial conditional intensity function is the failure rate function of the
original lifetime (time to first failure of the system), which is often a
continuous function of time, and is classified as constant, monotonic
increasing or decreasing, or a combination of these. Beyond the first failure,
the conditional intensity function is altered in accordance with the
rectifications performed following the first and all consequent failures.

Not all rectifications have the same effect on the system, and based on their
effect, they are categorized as either replacements or repairs with varying
degrees of effectiveness. In some cases, replacements can be viewed as extreme
repairs.

In this article, we suggest an approach to model the effect of rectifications
(here, repairs) for a system having a non-monotonic (here,
bathtub-shaped) failure rate function. We define repairs in terms of their
effect on the virtual age of the system.

The article is arranged as follows. In Section \ref{background} we discuss the
concepts mentioned above in more detail, and provide a brief review of existing
models relevant to our study. In Section \ref{model_formulation}, we describe
the repair model and provide model formulation. In Section \ref{example}, we
provide a numerical illustration of the proposed model. Finally, in Section
\ref{conclusion}, we conclude with a discussion of the proposed model and some
directions for future research.

\section{Background and Definitions}
\label{background}

\begin{figure}[htb!]
\begin{center}
\begin{tabular}{cc}
\includegraphics[scale=0.26,angle=270]{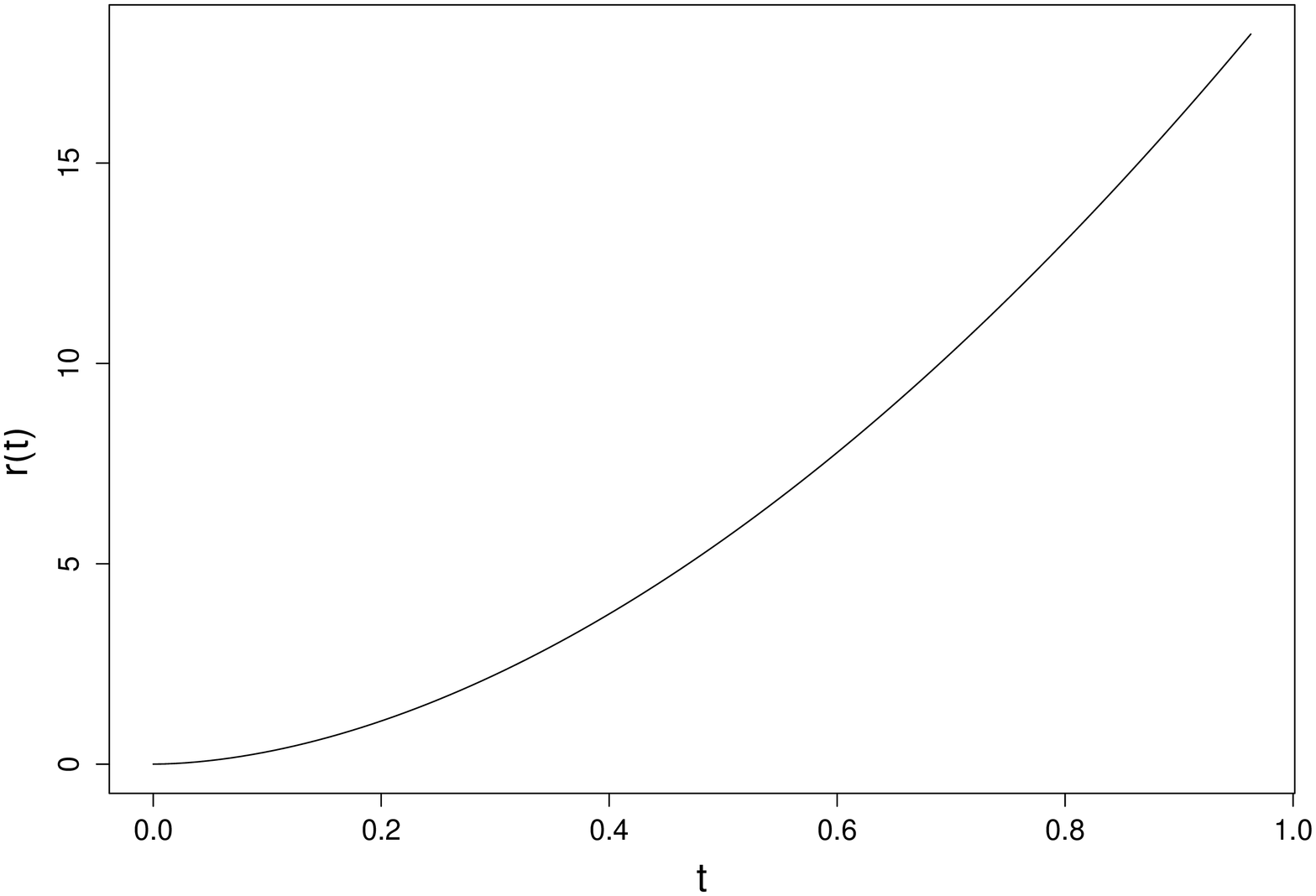} & \includegraphics[scale=0.26,angle=270]{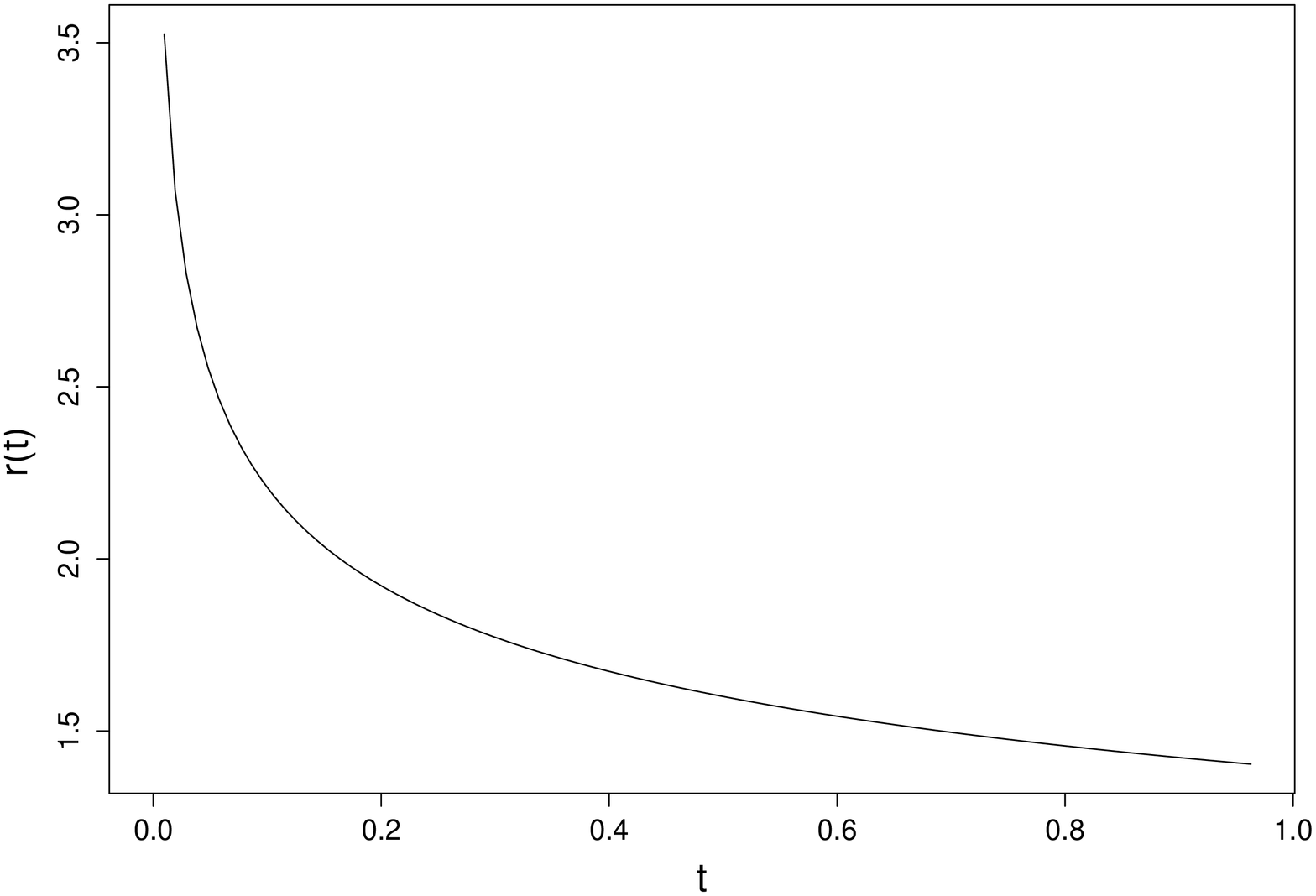}\\
{\footnotesize (i)} & {\footnotesize (ii)}\\
\includegraphics[scale=0.26,angle=270]{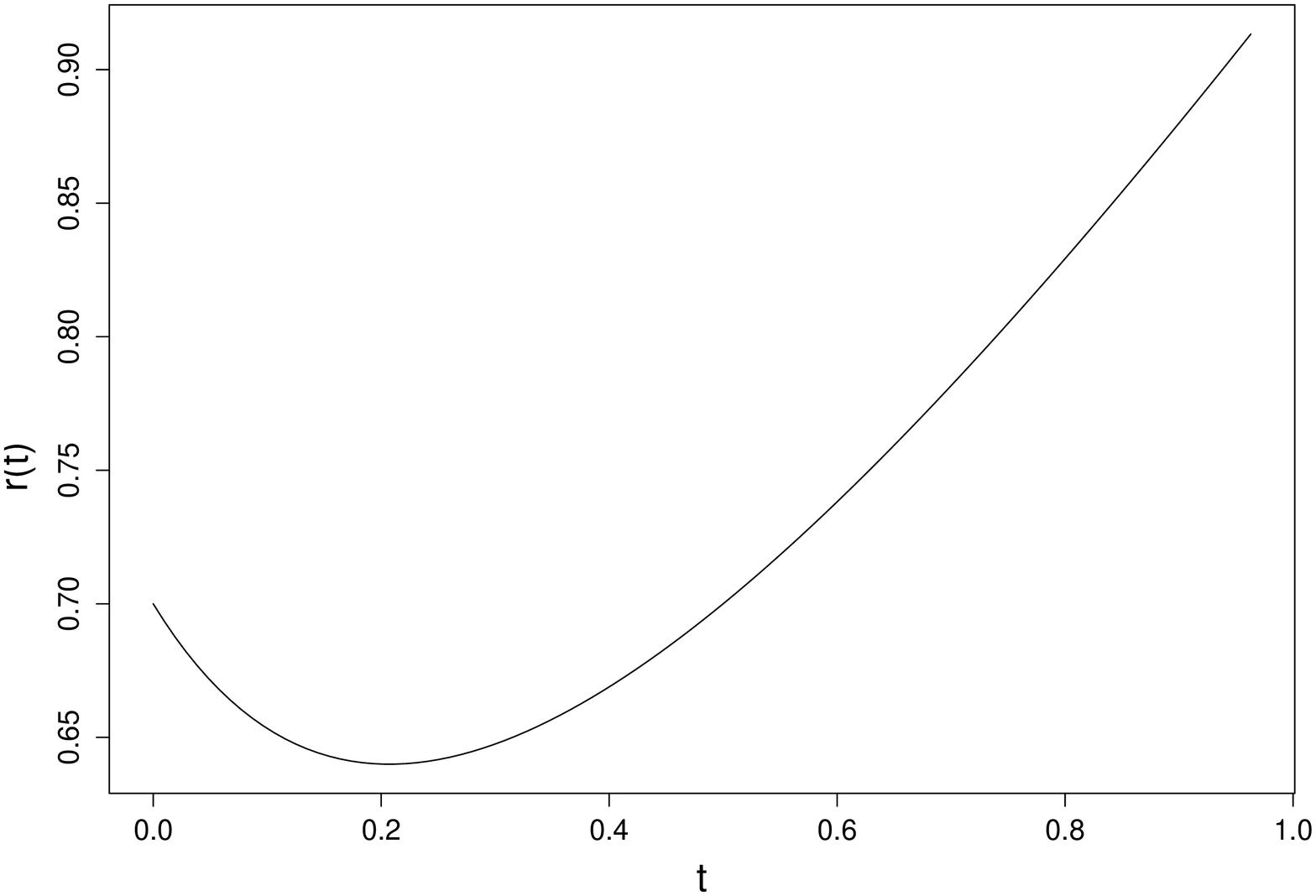} & \includegraphics[scale=0.26,angle=270]{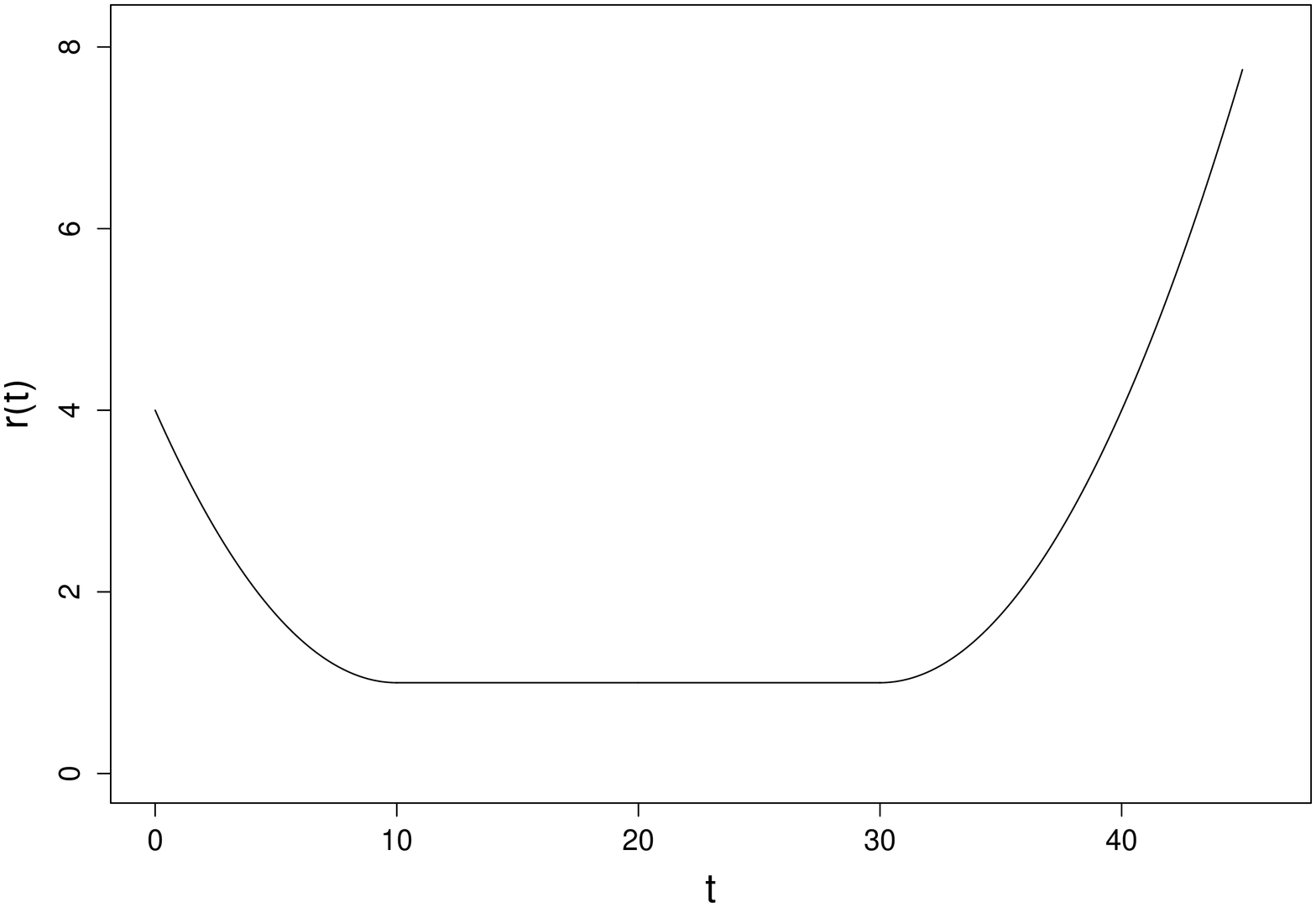}\\
{\footnotesize (iii)} & {\footnotesize (iv)}\\
\end{tabular}
\caption{Failure rate functions: (i) IFR;  (ii) DFR;  (iii) UFR;  (iv) BFR.}
\label{fr_functions}
\end{center}
\end{figure}

Let $\lambda_c(t)$ denote the conditional intensity function of the failure
process denoted by $\{N(t); t\geq0\}$. Then
\begin{equation*}
\lambda_c(t)=\lim_{dt\rightarrow 0} \frac{P\{N(t+dt)-N(t)=1\mid \mathcal{F}_t\}}{dt}\enspace,
\end{equation*}
where $N(t+dt)-N(t)$ is the number of failures in the interval $(t, t+dt]$, and
$\mathcal{F}_t=\{N(s); 0\leq s< t\}$ is the history of the process before time
$t$. The initial conditional intensity (or baseline intensity), denoted by
$\lambda_0(t)$, is the failure rate of the time to first failure, which is
\begin{displaymath}
\lambda_0(t) = r(t)= \lim_{dt\rightarrow 0} \frac{P\{N(t+dt)-N(t)=1\mid N(t)=0\}}{dt}\enspace.
\end{displaymath}
A failure rate or the corresponding distribution is categorized as: constant
failure rate (CFR) when $r(t)$ is constant over $t$; increasing failure rate
(IFR) when $r(t)$ is increasing in $t$; decreasing failure rate (DFR) when
$r(t)$ is decreasing in $t$; or some combination of these. For instance, the
bathtub-shaped failure rate (BFR) function which is initially decreasing, then
constant and finally increasing:
\begin{equation}
r(t) = \left\{ \begin{array}{ll}
r_1(t): r_1'(t) < 0 \enspace, & t \leq a_1 \\
r_2(t): r_2'(t) = 0 \enspace, &  a_1 < t \leq a_2 \\
r_3(t): r_3'(t) > 0 \enspace, & t > a_2\enspace,
\end{array}
\right.
\label{bfr}
\end{equation}
where $a_1$ and $a_2$ are the change points (points at which the the derivative
of the failure rate function changes sign) of the BFR function. The BFR
function is a generalization of the above categories; setting $a_1=a_2=0$
($a_1=a_2=\infty$ or $a_1=0$ and $a_2=\infty$), it becomes an increasing
(decreasing or constant) function. Setting $a_1=a_2=a$, we get a U-shaped
failure rate (UFR) function \cite{lai_stochastic_2006}; see Figure \ref{fr_functions}.

The quantity $r(t)\hspace{0.5mm} dt$ is the approximate probability that the system will
fail for the first time in $(t,t+dt]$. The distribution of the time to first
failure, denoted by $T_1$, is defined by the failure rate $r(t)$, but
subsequent failure times are affected by the type of rectification performed
after a failure.

Rectifications are broadly classified into repair and replacement. With
replacements, the system is replaced by a new, identical system upon failure.
The condition of the system following a replacement is therefore identical to a
new system. In this case, the failure process is modeled as a renewal process
with conditional intensity function $\lambda_c(t)=r(t-T_{N(t)})$ (where
$T_{N(t)}$ is the time of the last replacement (perfect repair)), and the expected number of
replacements is given by the renewal function
\cite{hokstad_failure_1997,hunter_renewal_1974,aven_stochastic_1998}. Replacements
apply when the system is beyond repair (or non-repairable) or when replacing
the system is more feasible than repairing it.

Repairs are characterized by their effectiveness, which is often referred to as
the \textit{degree of repair} -- the degree to which the functioning condition
of the system is restored following the repair. Based on this degree, repairs are
typically categorized as minimal, imperfect or perfect repair.

\textit{Minimal} repairs have no effect on the failure rate function; i.e., the
system immediately before and after the failure is in the same functioning
condition. Therefore, when all repairs are minimal, the failure process is a
non-stationary Poisson process with conditional intensity function
$\lambda_c(t)=r(t)$. The expected number of failures over an interval $[0,t)$
is then given by $E[N(t)]=\int\limits_0^t \lambda_c(s)\; ds$.

\textit{Perfect} repairs, for IFR functions, leave the repaired system in an
as-good-as-new working condition, which implies that a perfect repair is
equivalent to a replacement and can be modeled as one. This definition works
for systems having an IFR function, since the system at the start of its
lifetime has the lowest failure rate, i.e., it is at its best. However, if a
system has a failure rate function that is initially decreasing (e.g. BFR),
this definition does not hold, because a repair that leaves the system in an
as-good-as-new condition is actually worsening the system, since the failure
rate of the system at the start of its lifetime is higher than its failure rate
when it is working at its best. Therefore, we distinguish between replacement
and perfect repair, and describe a perfect repair as the best form of repair
(not taking into account improvements/upgrades). In other words, a perfect
repair is one that restores the functioning condition of the system to its
condition when it is performing at its best. This point of ideal performance is
at the start of the system's lifetime for a system with an IFR function, but
not for a system having an initially decreasing failure rate function.

\textit{Imperfect} repair, sometimes referred to as general repair, is any
repair that leaves the system in a functioning condition that is between the
functioning conditions following minimal and perfect repairs. For systems with
a IFR function, the definition of an imperfect repair has included the extremes
minimal repair and replacement (aka perfect repair). Here, imperfect repair
includes as its extremes minimal and perfect repairs, but not
replacements. Therefore, we distinguish between repair and replacement.
 
In most settings, the degree of repair is a variable with range $[0,1]$, where
a degree of zero corresponds to a minimal repair, a degree of one corresponds
to a \textit{perfect} repair, and a degree between these extremes corresponds
to an imperfect repair. Therefore, the higher the degree of repair, the bigger
the improvement in the functioning condition of the system.

Here, we do not consider repairs that can worsen the system or upgrades (or
improvements).

Many repair models have been suggested for systems having IFR functions. Some
common models are the virtual age models discussed in \citeN{kijima_results_1989}, \citeN{varnosafaderani_imperfect_2012} and \citeN{doyen_classes_2004};
and the intensity reduction models discussed in \citeN{lindqvist_statistical_1998}, \citeN{varnosafaderani_two_2012} and \citeN{doyen_classes_2004}. Models of repair in the case of BFR functions assume
that rectifications are either minimal or replacements. The virtual age models
for IFR functions have been applied to BFR functions; see \cite{dijoux_virtual_2009}, but due to the failure rate being initially
decreasing, repairs of degree greater than zero actually worsen the product.

In this article, we propose a new approach to modeling imperfect repairs for
systems having BFR functions which better suit the definitions of the types of
repair. The effects of repairs are described in terms of modifications to the
virtual age function of the system.

\section{Modeling the Effect of Repairs}
\label{model_formulation}

Let $T_i$ denote the time of the $i$th failure (also repair, since repairs are
immediate and instantaneous), and let $\delta_i$ denote the degree of the $i$th
repair. Also, let $A(t)$ denote the virtual age of the system at time $t$.

\begin{figure}[htb!]
\begin{center}
\includegraphics[scale=1.5]{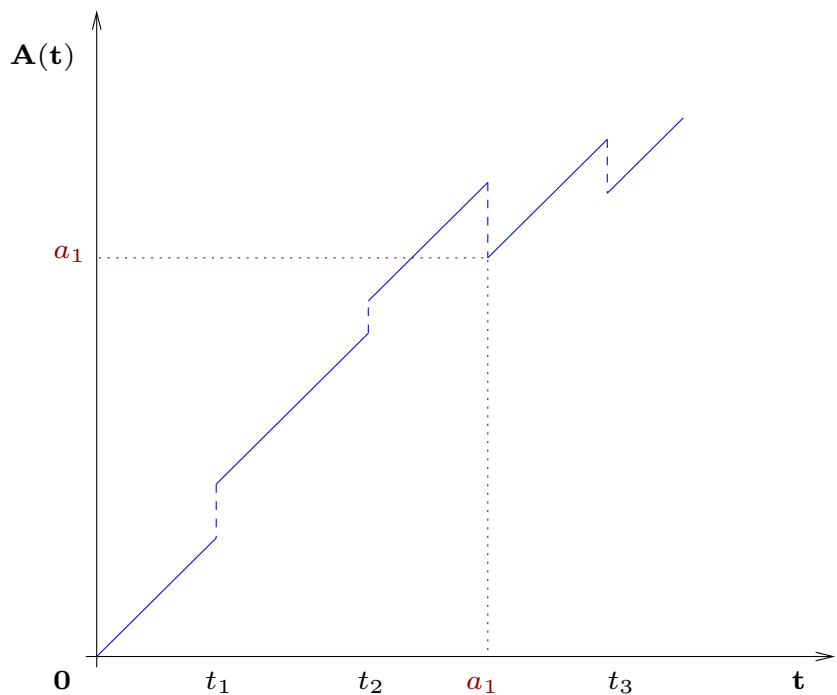}
\caption{Virtual age function following imperfect repairs of varying degree.}
\label{virtual_age}
\end{center}
\end{figure}

Based on the virtual age function of the system, we propose the following
repair model. The virtual age of the system at time $t$, is given by 
\begin{equation}
A(t) = \left\{ \begin{array}{ll}
  t + \sum\limits_{i=1}^{N(t^-)} \delta_i \; [a_1 - A(T_i)]\enspace, & \;\; t \leq a_1\\
  t - \sum\limits_{i=N(a_1^+)}^{N(t^-)} \delta_i \; [A(T_i) - a_1]\enspace, & \;\; t > a_1\\
 \end{array}\right.
\label{virtual_age_eq}
\end{equation}
where $A(T_i)$ is the virtual age of the system at the time of its $i$th failure. Before the first failure, when $N(t^-)=0$, the virtual age is simply $A(t)=t$; and immediately after $a_1$ and before the first failure in the useful life period, the virtual age of the system is again $A(t)=t$. See Figure \ref{virtual_age} for an illustration of the virtual age function for three failures occurring at times $t_1$, $t_2$ and $t_3$. 

With a failure rate function that is initially decreasing, the point of best
performance is not the start of the system's lifetime, but the point at which
the failure rate function is at its lowest. This point for a BFR function is the first change point $a_1$.

Therefore, according to this model, when the virtual age of the system at the time of the
$i$th failure is less than the first change point $a_1$, the effect of a repair
is modeled as an increase in the virtual age of the system, such that, a
perfect repair results in the virtual age being $a_1$. At $a_1$, the virtual age of the system is set to its calendar age, i.e. $A(a_1)=a_1$. This extends the useful life period of the system, which will decrease the probability of future failures. When the age of the system is greater than the first change point $a_1$, then the effect of a repair is a decrease in the virtual age of the system, such that, a perfect repair results in the virtual age being $a_1$. The point $a_1$ is the point of ideal performance, because it is the start of the useful life of the system, and the failure rate of the system at this point is at its lowest. 

\begin{figure}[htb!]
\begin{center}
\includegraphics[scale=0.55,angle=270]{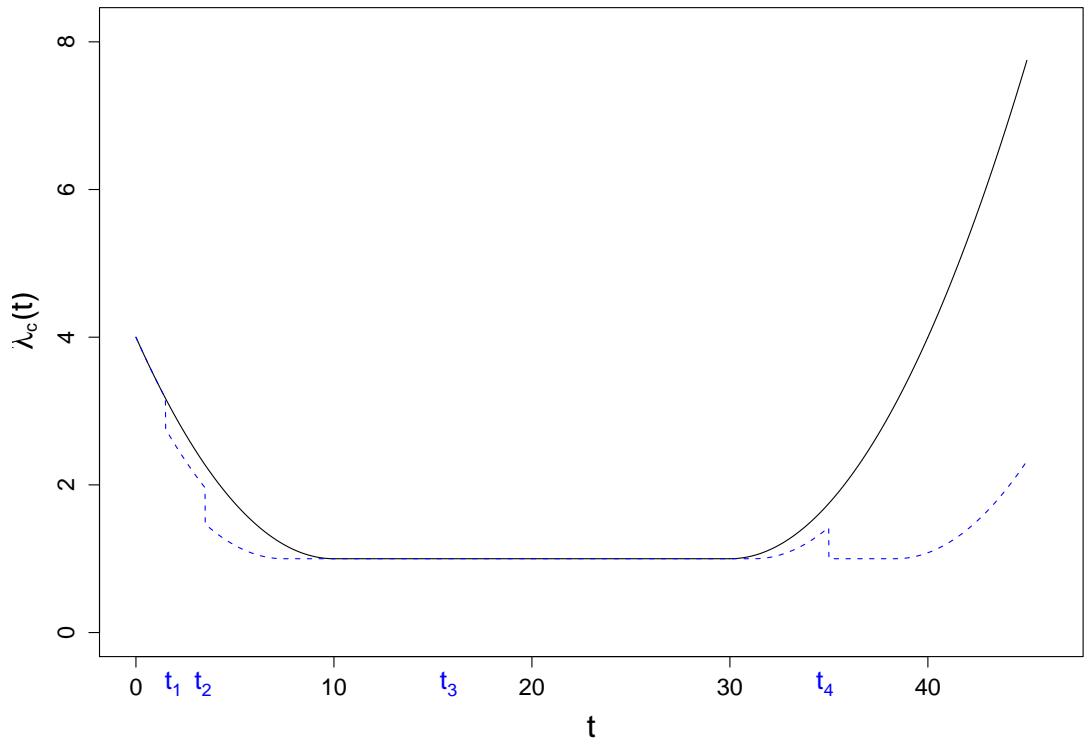}
\caption{Conditional intensity function following: four imperfect repairs of varying degree (dashed line); minimal repairs (solid line).}
\label{conditional_intensity}
\end{center}
\end{figure}

The conditional intensity function of the failure process is given by
\begin{displaymath}
\lambda_c(t) = \left\{ \begin{array}{ll}
 \lambda_0(t)\enspace, & t \leq T_1\\
 \lambda_0(A(t))\enspace, & t > T_1\\
 \end{array}\right.
\end{displaymath}
where $\lambda_0(.)$ is the baseline intensity (or failure rate) function. See Figure  \ref{conditional_intensity} for an illustration of this function following repairs of varying degree.

The repair model stays true to the definitions of the types of repair. A
perfect repair is the best form of repair, and should result in the system
performing at its best, which is in this case at $a_1$. A minimal repair,
should by definition leave the system in the same condition that it was prior
to failure, and here, the virtual age does not change following a minimal
repair. The effect of an imperfect repair should be between those of the
minimal and perfect repairs, and effectiveness of the repair should increase
with its degree. Here, as the degree of repair increases, so does the
effectiveness of the repair (which is reflected in the decrease in the
conditional intensity function of the process).

The assumption for this model is that the useful life period $(a_1,a_2]$ of the system is at least as long as the DFR period $(0,a_1]$, i.e. $a_2-a_1 \geq a_1$.   

\section{Numerical Illustration}
\label{example}

In this section, we provide a simple example that illustrates the proposed repair model. 

The baseline intensity function used in this example is
\begin{equation}
\lambda_0(t) = \left\{ \begin{array}{ll}\vspace{2mm}
\lambda + \alpha_1\; (a_1-t)^{\beta_1}\enspace, & t \leq a_1 \\ \vspace{2mm}
\lambda \enspace, &  a_1 < t \leq a_2 \\\vspace{2mm}
\lambda + \alpha_2\; (t-a_2)^{\beta_2} \enspace, & t > a_2\enspace,
\end{array}
\right.
\end{equation}
where $\lambda > 0$, $\beta_1, \beta_2 > 0$, $\beta_1\geq \beta_2$, and $\alpha_1,\alpha_2 > 0$. The parameter values are chosen to be $\lambda=1$,  $\alpha_1=0.6$, $\alpha_2=0.5$, $\beta_1=2.5$, and $\beta_2=2.8$, and the change points are chosen to be $a_1=4$ and $a_2=8$.

Since virtual age models for IFR functions have been frequently examined and the effect of repairs in this case is known, we limit our illustration to exploring the effect of repairs based on our virtual age model in the DFR phase. To do so, we select an arbitrary mission time $\tau$, and applying repairs of varying degree in the interval $[0,a_1)$, we compute the expected number of failures in $(0,\tau]$. Here, the mission time is chosen to be $\tau=10$.

The repairs performed are chosen according to the following strategy: the first repair in the interval $(0,a_1]$ is imperfect, and all other repairs are minimal.

Let $T_1$ denote the time of the first failure. The density function of $T_1$ in terms of the baseline intensity function is given by
\begin{equation}
f_1(t) = \lambda_0(t)\; e^{-\int\limits_0^t \lambda_0(s)\;ds} \enspace.
\label{ft1}
\end{equation}

The expected number of failures in the interval $[0,\tau)$, is then derived as follows:
\begin{displaymath}
\begin{aligned}
 E[N(\tau)] = & \int\limits_0^{a_1} \bigg[1 + \int\limits_{t_1}^{a_1} \lambda_0(s+\delta_1(a_1-t_1))\; ds\bigg] \; f_1(t_1)\; dt_1 + \int\limits_{a_1}^{\tau} \lambda_0(s)\; ds \enspace,\\
\end{aligned}
\end{displaymath}
where $\delta_1$ is the degree of the imperfect repair performed in $(0,a_1]$.

Tabulated in Table \ref{results} are the expected numbers of failures $E[N(10)]$ for degrees of repair $\delta_1 \in \{0.1,0.2,\dots,1.0\}$. 

\begin{table}[htb!]
\caption{Expected number of failures in the interval $[0,\tau)$ for various degrees of repair}
{\footnotesize
\begin{center}
 \begin{tabular}{|c|c|c|c|c|c|c|c|c|c|c|c|c|c|c|}
\hline
$\mathbf{\delta_1}$ & 0.0  &  0.1  &  0.2 & 0.3 & 0.4 & 0.5 & 0.6 & 0.7 & 0.8 & 0.9 & 1.0 \\
\hline
$E[N(10)]$ & 33.78 & 27.3 & 22.4 & 18.81 & 16.29 & 14.64 & 13.63 & 13.09 & 12.86 & 12.79 & 12.78 \\
\hline
 \end{tabular}
\end{center}
}
\label{results}
\end{table}

\begin{figure}[htb!]
\begin{center}
\includegraphics[scale=0.55,angle=270]{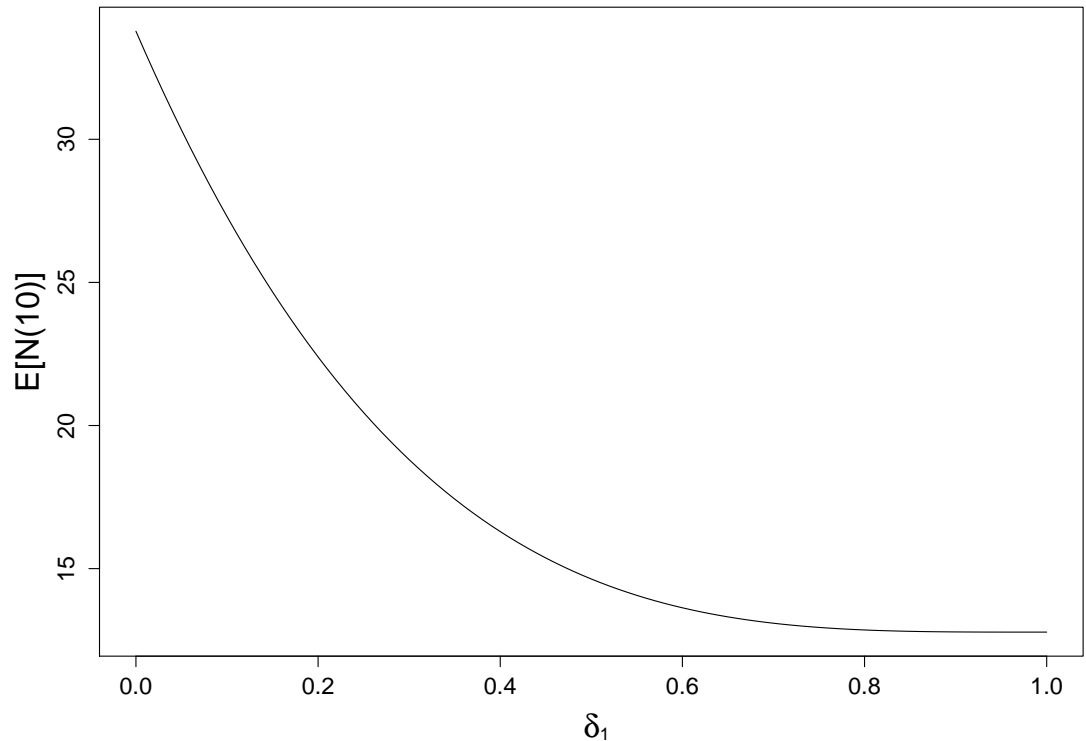}
\caption{Expected number of failures $E[N(10)]$ for $\delta_1 \in [0,1]$.}
\label{results_plot}
\end{center}
\end{figure}

Note that, according to the repair model, as the degree of repair increases, the expected number of failures decreases; also see Figure \ref{results_plot}.

\section{Conclusion}
\label{conclusion}

In this article, we proposed a new repair model for systems having a BFR function. The effect of repairs was modeled as a modification in the virtual age of the system following the repairs.

According to the proposed model (illustrated in Section \ref{example}), as the degree of any given repair increases (while others remain fixed), the expected number of failures decreases, since the reliability of the system is improved. 

Some possible future research directions are deriving virtual age models for systems with more than two change points and extension of these models to two dimensions.


\bibliographystyle{achicago}
\bibliography{references}

\end{document}